# 3DGauCIM: Accelerating Static/Dynamic 3D Gaussian Splatting via Digital CIM for High Frame Rate Real-Time Edge Rendering

Wei-Hsing Huang

Georgia Institute of Technology, whuang386@gatech.edu

Cheng-Jhih Shih

Georgia Institute of Technology, cshih63@gatech.edu

Jian-Wei Su

National Tsing Hua University, jwsu@gapp.nthu.edu.tw

Samuel Wade Wang

Georgia Institute of Technology, swang3068@gatech.edu

Vaidehi Garg

Georgia Institute of Technology, vgarg34@gatech.edu

Yuyao Kong

Georgia Institute of Technology, ykong62@gatech.edu

Jen-Chun Tien

National Tsing Hua University, jimytien01@gmail.com

Nealson Li

Georgia Institute of Technology, nealson@gatech.edu

Arijit Raychowdhury

Georgia Institute of Technology, arijit.raychowdhury@ece.gatech.edu

Meng-Fan Chang

National Tsing Hua University, mfchang@ee.nthu.edu.tw

Yingyan (Celine) Lin

Georgia Institute of Technology, celine.lin@gatech.edu

Shimeng Yu

Georgia Institute of Technology, shimeng.yu@ece.gatech.edu

Abstract—Dynamic 3D Gaussian splatting (3DGS) extends static 3DGS to render dynamic scenes, enabling AR/VR applications with moving objects. However, implementing dynamic 3DGS on edge devices faces challenges: (1) Loading all Gaussian parameters from DRAM for frustum culling incurs high energy costs. (2) Increased parameters for dynamic scenes elevate sorting latency and energy consumption. (3) Limited on-chip buffer capacity with higher parameters reduces buffer reuse, causing frequent DRAM access. (4) Dynamic 3DGS operations are not readily compatible with digital compute-in-memory (DCIM). These challenges hinder real-time performance and power efficiency on edge devices, leading to reduced battery life or requiring bulky batteries. To tackle these challenges, we propose algorithm-hardware co-design techniques. At the algorithmic level, we introduce three optimizations: (1) DRAM-access reduction frustum culling to lower DRAM access overhead, (2) Adaptive tile grouping to enhance on-chip buffer reuse, and (3) Adaptive interval initialization Bucket-Bitonic sort to reduce sorting latency. At the hardware level, we present a DCIM-friendly computation flow that is evaluated using the measured data from a 16nm DCIM prototype chip. Our experimental results on Large-Scale Real-World Static/Dynamic Datasets demonstrate the ability to achieve high frame rate real-time rendering exceeding 200 frame per second (FPS) with minimal power consumption—merely 0.28 W for static Large-Scale Real-World scenes and 0.63 W for dynamic Large-Scale Real-World scenes. This work successfully addresses the significant challenges of implementing static/dynamic 3DGS technology on resource-constrained edge devices.

**ACM Reference Format**

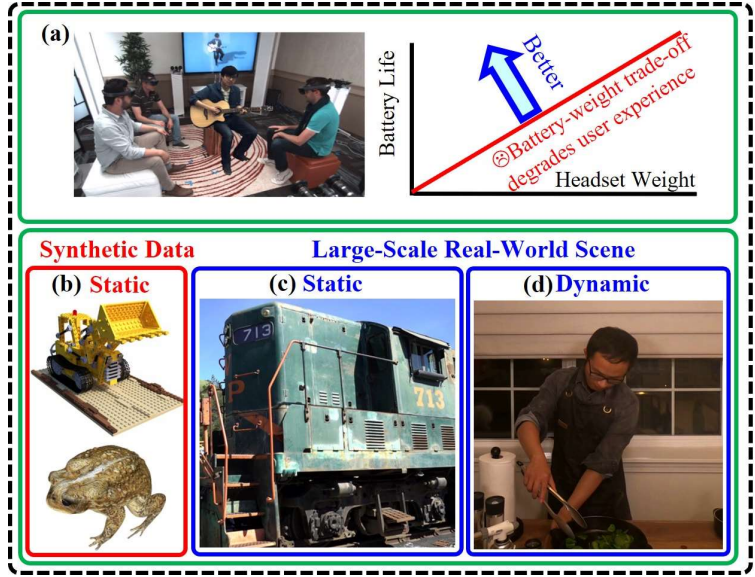

Fig. 1(a) Illustration of dynamic 3D Gaussian splatting applications [19] (b) Small-Scale synthetic datasets [24,25,26] (c) Large-Scale Real-World static scene [22] (d) Large-Scale Real-World dynamic scene [21].

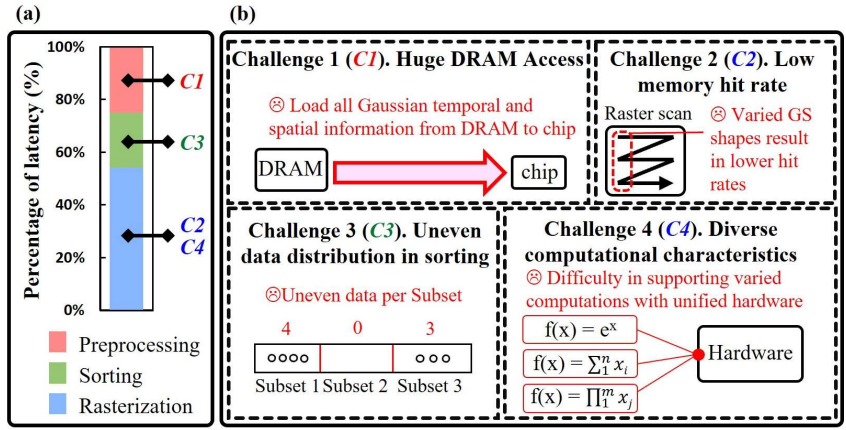

Fig. 2(a) Latency profiling of dynamic 3DGS (b) Challenges hindering low-power real time rendering edge applications.

# 1 INTRODUCTION

Augmented and Virtual Reality (AR/VR) [1] are advancing human-computer interaction paradigms, enabling immersive experiences across gaming, training, collaboration, and autonomous driving domains. The pursuit of enhanced realism necessitates real-time photorealistic rendering capabilities. 3D Gaussian Splatting (3DGS) [2] has emerged as a prominent technique to address this need, demonstrating significant potential in this domain. Dynamic 3D Gaussian Splatting (3DGS) [3] extends to dynamic scenes (Fig. 1(a)), integrating moving objects with static environments, essential for applications like virtual tourism and meetings. However, dynamic 3DGS faces challenges due to increased computational complexity and parameter count compared to static 3DGS. Scene types include Small-Scale synthetic datasets (Fig. 1(b)) [24-26] and Large-Scale Real-World scenes (Fig. 1(c, d)) [21, 22]. This work focuses on both static and dynamic Large-Scale Real-World scenes. Among end-to-end accelerators related to our work which support Large-Scale Real-World scenes, [4] only supports Large-Scale Real-World static scene but cannot achieve high frame rate real-time rendering (only 91.2 FPS) and lacks support for dynamic scenes. We profiled the Gaussian-splatting kernel [27] on Large-Scale Read-World dynamic scenes [21] using NVIDIA Nsight Systems [28] to obtain a detailed breakdown of latency. Profiling (Fig. 2(a)) highlights three key computational phases in dynamic 3DGS: preprocessing, sorting, and rasterization. Frustum culling dominates preprocessing time, a bottleneck significantly exacerbated by the temporal dimension compared to static 3DGS. Based on the profiling results, we identify four key challenges (Fig. 2(b)) that constitute performance bottlenecks within the preprocessing, sorting, and rasterization phases. Therefore, we propose an end-to-end algorithm-hardware co-design framework for static/dynamic 3DGS, which to our knowledge is the first to optimize Large-Scale Real-World dynamic scenarios. This framework achieves high frame rate real-time performance (>200 FPS) with exceptional efficiency (0.28W for static and 0.63W for dynamic scenes).

The primary contributions of this work are listed as follows:

- We propose DRAM-access reduction frustum culling (DR-FC) to address Challenge 1 presented in Fig. 2(b). This approach performs offline partitioning of the 3D scene volume into a coarse-grained grid structure. This coarse-grained grid enables efficient frustum culling without requiring DRAM access to complete Gaussian parameters when given a camera pose at time t. While effective for both static and dynamic scenes, DR-FC is particularly crucial for dynamic scenes where the temporal dimension substantially expands the parameter count. By eliminating loading a substantial number of out-of-frustum Gaussian parameters during preprocessing, this method significantly reduces the required costly DRAM read operations.
- We propose Adaptive Tile Grouping with the posteriori knowledge (ATG) to address Challenge 2 presented in Fig. 2(b). ATG tracks the spatial relationships between Gaussians and tiles during intersection testing. Based on these tracked relationships, the system dynamically optimizes tile grouping to maximize the on-chip buffer reuse efficiency and leverages the posteriori knowledge from the previous frame to enable efficient tile group generation for the current frame, achieving both low latency and low power consumption.
- In contrast to conventional partition-based sorting algorithms, where boundary point selection often results in unbalanced partition distributions, as illustrated in Challenge 3 of Fig. 2(b), we present Adaptive Interval Initialization Bucket-Bitonic Sort with the posteriori knowledge (AII-Sort) to address this data distribution challenge. AII-Sort leverages the frame-to-frame correlation of Gaussian Splatting scenes by utilizing boundary points from the previous frame to initialize the current frame's bucket intervals. This approach achieves balanced bucket distributions, enabling efficient sorting operations.
- As shown in Challenge 4 of Fig. 2(b), dynamic 3DGS requires diverse intensive computations. Instead of implementing customized hardware for each computation type, which increases system area and complexity, we propose a DCIM-friendly dynamic 3DGS dataflow (DD3D-Flow) that efficiently maps these diverse computations to the DCIM architecture. The statistics of measured DCIM chips [5] are used.

## 2 BACKGROUND

### 2.1 Dynamic Scenes Using Gaussian Splatting

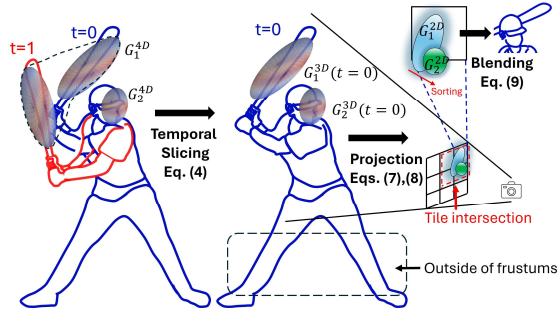

Fig. 3. The complete pipeline of dynamic 3D GS onto the pixel space.

Dynamic scene reconstruction poses challenges due to its varying spatial-temporal elements. While traditional frame-by-frame approaches [6] process each frame independently, they fail to utilize temporal continuity. MLP-based methods offer alternatives: [9] incorporates temporal data directly, while [7] employs a spatial-temporal encoder before MLP processing. However, these MLP-based approaches suffer from decreased rendering efficiency. The 4D Gaussian Splatting (4DGS) technique [8, 10] represents scenes using 4D Gaussian primitives in space (x,y,z) and time (t), enabling smooth dynamic changes and end-to-end training. We adopt 4DGS for its scalability and continuous temporal representation, which efficiently handles complex dynamics while maintaining high-quality rendering. Throughout this paper, our discussion of Dynamic 3DGS refers to this 4DGS implementation, with the complete workflow shown in Fig. 3.

Refactor: Given $\mu \in R^n, \Sigma \in R^{n \times n}$, write:
$$\mathcal{G}(x; \mu, \Sigma) = exp(-(x-\mu)^T \Sigma^{-1}(x-\mu)/2) \tag{1}$$

4D Gaussian Representation: The 4D Gaussian representation is defined as follows: $x \in R^3, t \in R$, write:
$$G^{4D}((x,t)) = \mathcal{G}((x,t); \mu^{4D}, \Sigma^{4D}) \tag{2}$$

where $\mu^{4D} = (\mu_x, \mu_y, \mu_z, \mu_t), \Sigma^{4D}$ are the mean vector and the covariance matrix of the 4D Gaussian, with $\Sigma^{4D}$ admitting a spectral decomposition:
$$\Sigma^{4D} = USS^T U^T \tag{3}$$

with $U$ orthogonal, $S$ diagonal, as rotation and scaling.

4D to 3D Temporal Slicing: To render an image at a specific time t, we project the 4D Gaussian $G^{4D}$ down to a conditional 3D Gaussian $G^{3D}$ through temporal slicing:
$$G^{3D}(x,t) = \mathcal{G}(t; \mu_t, \lambda^{-1}) \cdot \mathcal{G}(x; \mu^{3D|t}, \Sigma^{3D|t}) \tag{4}$$

Where $\lambda = {\Sigma^{4D}_{4,4}}^{-1} > 0$ is a temporal decay parameter, $\mu^{3D|t}, \Sigma^{3D|t}$ are the conditional mean, covariance for the 3D slice, given by:
$$\mu^{3D|t} = \mu^{4D}_{1:3} + \Sigma^{4D}_{1:3,4} \cdot \lambda \cdot (t - \mu_t) \tag{5}$$
$$\Sigma^{3D|t} = \Sigma^{4D}_{1:3,1:3} - \Sigma^{4D}_{1:3,4} \cdot \lambda \cdot \Sigma^{4D}_{4,1:3} \tag{6}$$

3D to 2D Projection and Image Rendering: The resulting 3D Gaussian $G^{3D}(x,t)$ is projected onto the 2D image plane by transforming its mean and covariance accordingly:
$$\mu^{2D} = \text{Proj}(\mu^{3D|t} E, K)_{1:2} \tag{7}$$
$$\Sigma^{2D} = (JW\Sigma^{3D|t}W^T J^T)_{1:2,1:2} \tag{8}$$

where $Proj(\cdot\ |E,K)$ performs the projection using camera extrinsics $E$ and intrinsics $K$, $J$ is the Jacobian matrix of the projection, and $W$ is the viewing transformation. Finally, the image at pixel $(u,v)$ is rendered by blending the visible 2D Gaussian splats:

$$I(u,v,t) = \sum_{i=1}^{N} \alpha_i c_i(d) \prod_{j=1}^{i-1}(1-\alpha_j) \tag{9}$$

where $d$ is the current viewing direction, $i,j$ represent indices of Gaussians in the scene, $N$ is the total number of Gaussians, and $o_i, c_i(d), \prod_{j=1}^{i-1}(1-\alpha_j)$ are the opacity, view dependent color, transmittance of the $i$-th Gaussian, where $\alpha_i$ is given by:

$$\alpha_i = o_i \cdot \mathcal{G}(t;\mu_{t,i},\lambda_i^{-1}) \cdot \mathcal{G}\left((u,v);\mu_i^{2D},\Sigma_i^{2D}\right) \tag{10}$$

$c_i(d)$ is computed by Spherical Harmonics (SH) function with SH parameters [2].

While mapping equation (10) into our hardware, we merge $\mathcal{G}(t;\mu_{t,i},\lambda_i^{-1}) \cdot \mathcal{G}((u,v);\mu^{2D,i},\Sigma^{2D,i})$ together to $P_i(u,v,t)$ using one exp function for hardware efficiency. In the following, we refer to the SH parameter, opacity as Gaussian's feature parameters.

## 2.2 User Behavior Analysis in Head Movement

User behavior analysis in head movement reveals fundamental patterns in human-platform interaction. [11] examined head movement patterns across head-mounted displays, smartphones, and PCs using viewport data from 275 users over 156 hours. The study revealed that frame-to-frame angular movements remain small across all interaction methods, with angles confined to modest ranges in both latitude and longitude. This behavior exhibits temporal coherence, with limited angular displacement between consecutive frames. The inherent frame-to-frame correlation characteristics motivate our optimization approach for the 3DGauCIM acceleration framework.

## 2.3 Compute-in-memory

Compute-in-Memory (CIM) architectures employ SRAM, eDRAM, and emerging non-volatile memories like RRAM for analog computation via column-wise current/charge summation [12,13]. While effectively addressing the memory wall, analog CIM suffers from accuracy degradation due to process variations and scaling limitations in advanced nodes. In contrast, DCIM [14] offers superior scalability in advanced nodes, preserving input parallelism and accuracy. 4T Gain Cell-based DCIM [5] further enhances density and efficiency due to its reduced transistor count, accelerating MAC operations with lower energy, suitable for low-power, real-time 3DGS. Therefore, this work adopts 4T Gain Cell-based DCIM [5] as the computing unit. However, efficient mapping of diverse, intensive computations required by dynamic 3DGS onto DCIM is currently lacking. In Section 3.4, we propose DCIM friendly dynamic 3DGS dataflow to address this challenge.

## 3 PROPOSED TOP-DOWN HW-SW CO-OPTIMIZATION

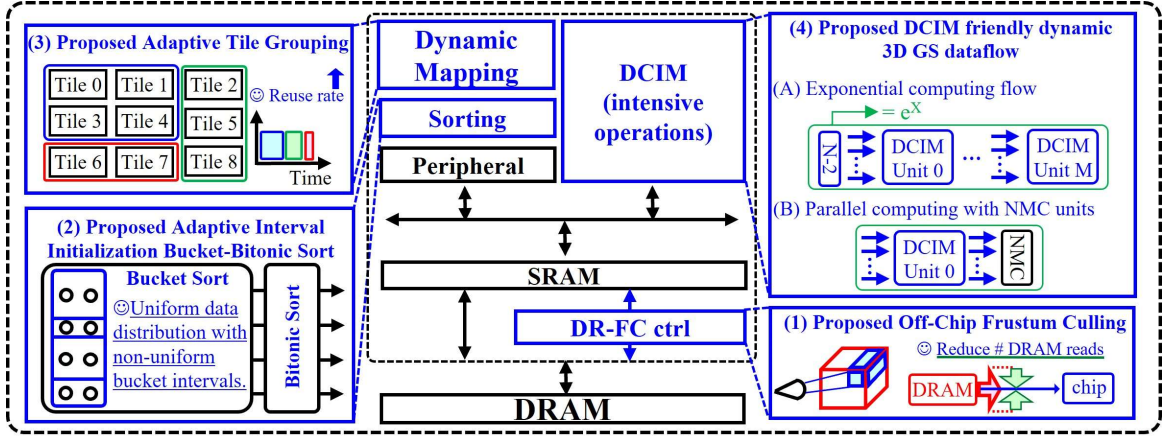

Fig. 4. Overall dataflow and hardware architecture of dynamic 3DGS accelerator.

Fig. 4 illustrates the overall accelerator architecture. The proposed dynamic 3DGS acceleration framework introduces four key features: (1) DRAM-access reduction frustum culling, which effectively reduces DRAM read operations; (2) Adaptive Interval Initialization Bucket-Bitonic Sort with the posteriori knowledge, enabling self-adjusting bucket intervals to achieve Near-Perfect Interval initialization for each bucket in sorting stage; (3) Adaptive tile grouping with the posteriori knowledge, designed to improve data reuse efficiency in on-chip SRAM buffer in blending stage; (4) DCIM friendly dynamic 3DGS dataflow for blending stage.

### 3.1 DRAM-access reduction frustum culling

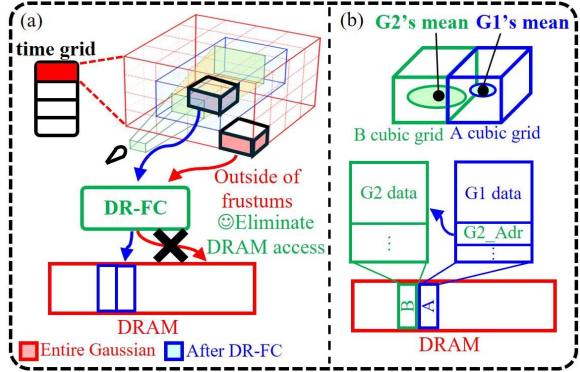

Fig. 5. Offline coarse-grain DRAM-access reduction frustum culling.

As illustrated in Challenge 1 of Fig. 2(b), traditional frustum culling [4] involves fetching all Gaussians from off-chip DRAM to exclude Gaussians outside the viewing frustum. This process leads to significant computational overhead, high energy consumption, and increased latency, especially in Large-Scale Real-World dynamic scenes. To address these inefficiencies, we propose a DRAM-access reduction frustum culling (DR-FC) method, as shown in Fig. 5(a). In dynamic 3DGS, each Gaussian parameter includes a position and temporal mean, along with covariance, as defined in equation (2).

Our approach uses a two-stage partitioning strategy: first, distributing Gaussians into coarse 1D temporal grids based on temporal means, then partitioning each 1D grid into coarse cubic grids by position means. This offline partitioning only loads grid information to an on-chip buffer during initialization. Since Gaussians within each cubic grid are stored contiguously in DRAM, our method only requires the buffer to store the start and end addresses of Gaussians for each grid, making it memory efficient. Given a camera pose and time t, our method can identify out-of-frustum cubic grids without accessing DRAM, thereby significantly reducing DRAM accesses. Next, we optimize Gaussian placement within these grids. Although our strategy is described with cubic grids, it can be extended to 1D grids as well. To enhance DRAM burst read efficiency, Gaussians within each cubic grid are stored contiguously in DRAM. Initially, each Gaussian is placed in its central cubic grid based on its mean. However, due to their covariance, some Gaussians span multiple cubic grids. To handle this, we store the complete Gaussian data in the central grid, while neighboring grids only hold pointers to this data, minimizing memory overhead, as shown in Fig. 5(b). Additionally, Gaussians spanning adjacent cubic grids are stored contiguously in memory within their central grid, for efficient access when referenced from neighboring grids. When both central and adjacent cubic grids are visible simultaneously, redundant DRAM accesses could increase energy consumption and latency. To mitigate this, we integrate DR-FC with an efficient memory access strategy. During DR-FC, the controller records all grids within the viewing frustum. If a Gaussian referenced in an adjacent grid is already scheduled for access through its central grid, the duplicate reference is skipped, avoiding redundant DRAM operations. This coarse-grain frustum culling before initiating DRAM operations markedly reduces memory access, improving overall system efficiency.

### 3.2 Adaptive Interval Initialization Bucket-Bitonic Sort with the posteriori knowledge

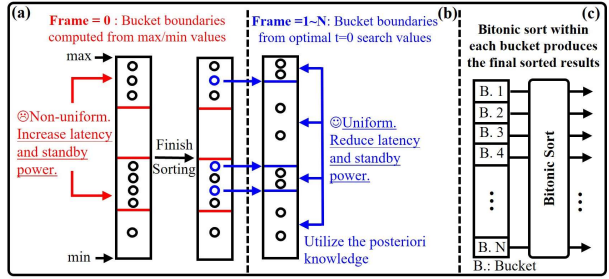

Fig. 6. Adaptive Interval Initialization Bucket-Bitonic Sort with the posteriori knowledge.

In the sorting stage for edge devices, the selection of an appropriate sorting algorithm is crucial. While hardware-intensive parallel sorting algorithms offer high performance at significant hardware cost, simpler algorithms like Insertion Sort and Bubble Sort [15] are hardware-efficient but performance-limited. Although Bucket Sort [16] provides a middle ground, its effectiveness relies heavily on uniform data distribution. Therefore, we propose Adaptive Interval Initialization Bucket-Bitonic Sort with the posteriori knowledge (AII-Sort), leveraging frame continuity to enable subsequent frames to retain bucket ranges from preceding frames. This propagation mechanism facilitates a near-uniform distribution of data within each bucket, thereby achieving the near optimal performance of Bucket Sort with an amortized $O(N)$ time complexity, as illustrated in Fig. 6(a-b). AII-Sort operates in two phases:

*A. Phase One: Initial Frame Processing.*

For frame 0, AII-Sort first determines the maximum and minimum depth values across all Gaussians during preprocessing. In the sorting stage, given N predefined buckets, the algorithm divides the range between these maximum and minimum depth values into N uniform segments, establishing initial bucket boundaries.

*B. Phase Two: Subsequent Frame Processing.*

For frames 1 to N, AII-Sort leverages frame-to-frame correlation by initializing the current frame's bucket sort with the previous frame's sorted bucket ranges. This approach facilitates a nearly uniform distribution of data across each bucket, thereby enabling the bucket sort algorithm to achieve performance approximating the theoretical optimum with an average time complexity of O(N). In addition, this approach avoids computing the maximum and minimum depth values across all Gaussian components during the Phase One preprocessing stage, significantly enhancing sorting efficiency.

As illustrated in Fig. 6(c), after allocating each data point to its bucket, Bitonic Sort [17,18] is applied to sort the data, obtaining the final result. Since Gaussians frequently splat to adjacent tiles, recording fine-grained bucket intervals for individual tiles is unnecessary. Instead, we group adjacent tiles into Tile Blocks and store the average bucket interval value for each tile group.

### 3.3 Adaptive tile grouping with the posteriori knowledge

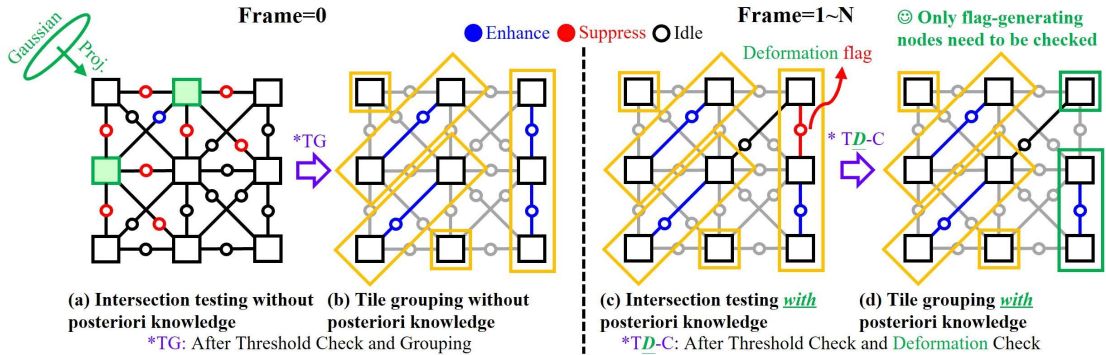

Fig. 7. Adaptive tile grouping with the posteriori knowledge.

During rendering, Gaussian elements frequently intersect multiple tiles. The conventional raster scan approach compromises Gaussian reusability in on-chip SRAM buffer, as illustrated in Challenge 2 of Fig. 2(b). For example, a vertical Gaussian spanning multiple tiles can result in decreased buffer reuse efficiency and increased DRAM reloads, leading to higher energy consumption and latency. The proposed Adaptive Tile Grouping with the posteriori knowledge (ATG) enhances buffer efficiency by recording and utilizing Gaussian-tile intersection patterns. ATG operates in two phases:

*A. Phase one: Without the posteriori knowledge.*

At frame=0, the system performs intersection testing by tracking the connection strength between tiles based on Gaussian overlapping patterns. As illustrated in Fig. 7(a), when a Gaussian intersects with two tiles, the connection strength at their shared boundary increases, while the connection strength at other boundaries decreases. This approach enhances Gaussian-tile intersection features. Based on a user-defined threshold, connections below this threshold are removed, and connected tiles are grouped using the Union-Find based algorithm, optimizing on-chip buffer data reuse, as illustrated in Fig. 7(b).

*B. Phase two: With the posteriori knowledge.*

From frames 1 to N, leveraging frame-to-frame correlation where Gaussian-tile intersections remain relatively stable, the system initializes using previous grouping information. As shown in Fig. 7(c), only boundary conditions that differ from the previous frame's grouping trigger a deformation flag. The system then selectively regroups only the flagged regions, eliminating the need for complete tile regrouping using the Union-Find based algorithm, thus achieving optimal tile grouping with low latency and energy, as shown in Fig. 7(d).

Implementation considerations:

I. As discussed in Section 3.2, we group neighboring tiles into Tile Blocks and track block-level connections for efficient processing.
II. II. During execution, ATG records the K highest and K lowest connectivity strengths within each tile and determines their respective median values as the upper and lower bounds. The final threshold is calculated using equation (11):

$$threshold = (upper - lower) \times user\_defined\_threshold + lower \qquad (11)$$

III. We co-optimize our caching strategy with the AII-Sort algorithm. For blending operations, the SRAM buffer is partitioned into N equal segments, where N corresponds to the number of buckets in AII-Sort. Gaussian parameters loaded from DRAM are stored in these N segments based on their depth values. This depth-based partitioning allows for efficient parameter retrieval during blending by first narrowing down the search range according to the Gaussian depth interval. Subsequently, a 2-way associative cache lookup is performed within the selected segment to retrieve the specific parameters.

## 3.4 DCIM friendly dynamic 3DGS dataflow

As in Challenge 4 of Fig. 2(b), Dynamic 3DGS operations involve diverse computational tasks incompatible with DCIM architectures. We propose a DCIM-friendly Dynamic 3DGS dataflow (DD3D-Flow) to address this limitation. One bottleneck in Gaussian computation is the evaluation of exponential functions. Our proposed DD3D-Flow, illustrated in Fig. 8(a), efficiently maps exponential functions to DCIM for low-power computation through two phases. While equation (1) shows the complete Gaussian formula, we use $e^x$ as a representative example, with conclusions applicable to the full equation. It is important to note that static 3DGS can be considered a simplified case of dynamic 3DGS. Therefore, the DD3D-Flow presented in Section 3.4, although described in the context of dynamic 3DGS, is also applicable to static 3DGS.

### A. Phase One: Base Conversion.

As shown in Fig. 8(a), we transform the non-DCIM-compatible $e^x$ into $2^{\frac{x}{\ln 2}}$, where ln2 is fused offline into dynamic 3DGS parameters, eliminating on-chip computational overhead. This results in $2^{x'}$, where x' maintains floating-point format.

### B. Phase Two: Sign-Integer-Fraction Decouple.

The process (SIF), shown in Fig. 8(a), decomposes x' into integer and fractional components, resulting $2^{int} * 2^{frac}$. For negative values, a two's complement operation is performed on the fractional component, and $2^{-1}$ is incorporated into the $2^{int}$ term. The $2^{int}$ term requires only shift operations rather than costly multiplications, while $2^{frac}$ is computed using a look-up-table (LUT)-based DCIM flow. Our experiments demonstrate that a 12-bit precision fractional component maintains Peak Signal-to-Noise Ratio (PSNR) without degradation. Implementation uses a 12-bit LUT divided into four segments, each requiring 8 LUT values in DCIM. After $2^{int}$ shift operations and floating-point format conversion, the result processes through four cascaded DCIM stages for final output generation. This two-phase approach enables efficient, pipelined computation of base-2 exponential functions on DCIM architecture.

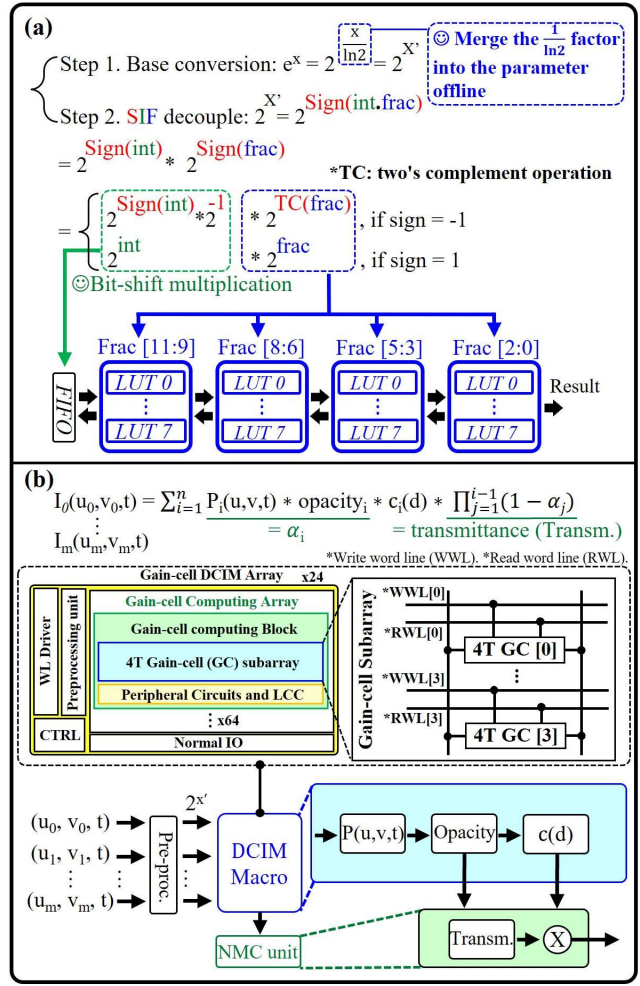

Fig. 8. DCIM [5] friendly dynamic 3D Gaussian Splatting dataflow.

Fig. 8(b) illustrates the DCIM Macro and our implementation of the blending operation presented in equation (9). The DCIM macro consists of 24 Gain-cell DCIM Arrays, where each array is composed of 64 Gain-cell computing blocks to enable parallel computation. Each Gain-cell computing block comprises a 64 bits gain-cell matrix integrated with a local computing cell (LCC). The DCIM stores three distinct types of data: (1) LUT for exponential computation, (2) opacity values, and (3) $c_i(d)$, view dependent RGB values, which are pre-computed by DCIM using Spherical Harmonics in the preprocessing stage. Multiple pixels are processed in parallel through the peripheral circuits for preprocessing before entering DCIM for massive parallel computation. Near-memory computing (NMC) units, positioned at the DCIM periphery, receive α values from DCIM and locally accumulate the transmittance. The final output is generated by combining the DCIM-computed α and RGB values with the locally accumulated transmittance in the NMC units.

## 4 EVALUATION RESULTS

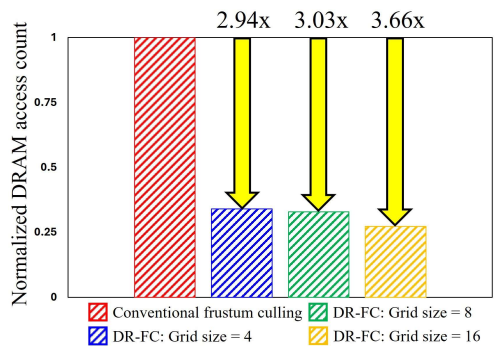

Fig. 9. Comparison of normalized DRAM access count between proposed DR-FC and conventional method across different grid sizes.

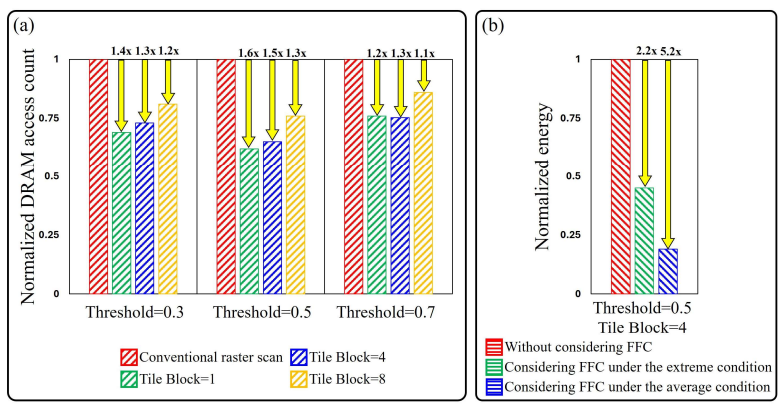

Fig. 10: (a) Comparison of normalized DRAM access count between ATG and conventional method. (b) Comparison of normalized energy using ATG without frame-to-frame correlation (FFC), with FFC in extreme conditions, and with FFC in average conditions.

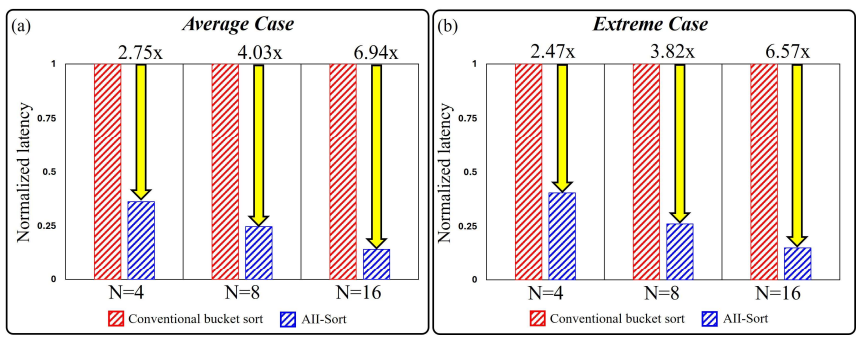

Fig. 11. Comparison of normalized sorting latency between proposed AII-Sort and conventional method in average and extreme conditions.

|  | 3DGauCIM | Jetson Orin[23] | 3DGauCIM | GSCore [4] |
| --- | --- | --- | --- | --- |
| Scene | **Real world Dynamic** Scene [21] | **Real world Dynamic** Scene [21] | **Real world Static** Scene [22] | **Real world Static** Scene [22] |
| Area (mm$^2$) | 4.13 | - | 1.81 | 3.95 |
| Power (W) | 0.63 | 15 | 0.28 | 0.87 |
| FPS | 211 | 31 | 214 | 91.2 |
| PSNR | 31.4 | 31.64 | 24.74 | 24.26 |
| SRAM buffer | 256KB | - | 256KB | 272KB |
| DCIM | 144KB | - | 48KB | - |
| Technology | 16nm | 8nm | 16nm | 28nm |

Table I: Comparison table of the proposed 3DGauCIM with GSCore and GPU in real-world Dynamic and Static Scenes.

In 3DGauCIM, we implement the system hardware using RTL, and perform synthesis, APR, and post-simulation using a commercial foundry's 16nm PDK. For DCIM, the statistics of measured TSMC's 16nm DCIM prototype chips [5] are used. Regarding DRAM, we adopt LPDDR5 and employ Ramulator 2.0 [20] for performance estimation. For the subsequent analysis, the numerical precision is set to FP16 format. For the analysis in Sections 4.A, 4.B, and 4.C, the Large-Scale Real-World dynamic scene [21] is used. To ensure analysis stability and accuracy, the performance metrics are averaged across all dynamic scenes in the dataset. In Section 4.D, we evaluate 3DGauCIM's performance using both Large-Scale Real-World dynamic scenes [21] and Large-Scale Real-World static scenes [22]. The results show that 3DGauCIM achieves excellent performance across static and dynamic scenarios.

*A. DRAM-access reduction frustum culling (DR-FC)*

In theory, finer grid partitioning enables more precise elimination of out-of-frustum grids without DRAM access, thereby maximizing DRAM access reduction. However, as discussed in Section 3.1, this approach increases the required on-chip buffer data storage. To quantitatively analyze the trade-off between additional on-chip buffer storage requirements and reduced DRAM access frequency across different grid granularities, we incrementally increased the number of grids. We compared DR-FC with conventional frustum culling methods that require loading all Gaussian data from DRAM. In Fig. 9, the grid number represents both the depth of 1D time grids and the dimensions (length, width, and height) of cubic grids. Fig. 9 demonstrates the effectiveness of our method: as grid numbers increased from 4 to 16, our approach showed significant improvements over conventional methods, achieving DRAM access reduction from 2.94x to 3.66x. This improvement is attributed to our method's ability to perform coarse-grained frustum culling without DRAM access, effectively eliminating a large number of out-of-frustum Gaussians.

*B. Adaptive tile grouping with the posteriori knowledge (ATG)*

To quantify the benefits of ATG, we compare DRAM access counts between conventional raster scan and ATG approaches. As discussed in Section 3.3, the user-defined threshold and Tile Blocks parameters significantly affect ATG's reuse efficiency. We conducted experiments varying the user-defined threshold from 0.3 to 0.7 and Tile Blocks from 1 to 8. Based on [11]'s analysis discussed in Section 2.2, screen-viewing users exhibit relatively limited angular movement, with median speeds of 14.8°/second (latitude) and 27.6°/second (longitude). We adopted these values for average condition analysis. Following [11], we set 180°/second as the maximum speed for both latitude and longitude to evaluate ATG under extreme conditions. Fig. 10(a) shows optimal performance at threshold = 0.5 and Tile Blocks = 1, achieving a 1.6x reduction in DRAM access count. A threshold of 0.3 degrades performance by forcing grouping of weakly correlated tiles within limited buffer capacity. Conversely, a threshold of 0.7 causes ATG to overlook many strongly correlated tile connections. While the configuration of threshold = 0.5 and Tile Blocks = 1 yields optimal performance, it increases memory overhead. Balancing this trade-off between memory overhead and DRAM access count reduction, we selected threshold = 0.5 and Tile Blocks = 4 for subsequent performance analysis. Fig. 10(b) analyzes both average and extreme conditions under this configuration. Considering frame-to-frame correlation, ATG achieves a 5.2x energy reduction in average conditions and a 2.2x reduction even in extreme conditions, demonstrating significant improvement in energy efficiency.

*C. Adaptive Interval Initialization Bucket-Bitonic Sort with the posteriori knowledge (AII-Sort)*

To evaluate AII-Sort's effectiveness, we compared the latency of AII-Sort against conventional Bucket-Bitonic sort. The Tile Blocks were configured to their optimal value of 4, as determined in Section 4.B, and tested under both average and extreme conditions. The traditional approach divides the Gaussian depth range into N uniform partitions between minimum and maximum values, as described in Section 3.2. We conducted experiments with N=4, 8, and 16. Fig. 11 shows that under average conditions, AII-Sort reduces latency by 2.75x to 6.94x as N increases from 4 to 16. Under extreme conditions, it maintains 2.47x to 6.57x latency reduction, demonstrating its robustness.

*D. DCIM friendly dynamic 3DGS dataflow (DD3D-Flow)*

The overall performance of 3DGauCIM is presented in Table I. Our evaluation is based on TSMC's 16nm DCIM chip measurement results [5], with all digital circuit modules verified through 16nm SPICE simulation for functionality. The optimal configuration demonstrates the following trade-offs: For DR-FC, we select a grid size of 4. While finer grid granularity could filter more Gaussians without DRAM access, the chosen configuration achieves a 2.94x reduction in DRAM access, as analyzed in Section 4.1, offering an optimal trade-off between memory overhead and DRAM access reduction. We configure Tile Blocks to 4 for both ATG and AII-Sort operations. The threshold for ATG is set to 0.5, and N is set to 8 for AII-Sort. As shown in Table I, the proposed DD3D-Flow, incorporating all our optimization schemes, achieves low power consumption of 0.28 W and real-time rendering at 203 FPS in Large-Scale Real-World static scenes. In more complex Large-Scale Real-World dynamic scenes, it maintains efficient performance with 0.63 W power consumption while delivering 211 FPS real-time rendering. These results demonstrate that our 3DGauCIM algorithm-hardware co-design framework achieves both low power consumption and real-time rendering capabilities across both Large-Scale Real-World static and complex dynamic scenarios. Table I further presents a comprehensive comparison between the proposed 3DGauCIM and GSCore [4]. While GSCore achieves 200 FPS on Small-Scale synthetic datasets [24,25], its performance decreases to approximately 100 FPS when processing Large-Scale Real-World static scenes [22]. In contrast, 3DGauCIM maintains a consistent 200 FPS performance across Large-Scale Real-World static and dynamic scenes. To the best of our knowledge, this work represents the first end-to-end accelerator for dynamic 3DGS. Therefore, for dynamic scene comparisons, we benchmark our solution against edge GPUs.

## 5 CONCLUSION

This paper introduces a novel 3DGauCIM framework that combines algorithm-hardware co-design for real-time rendering on power-constrained edge devices. In Large-Scale Real-World Datasets, our results demonstrate high frame rate real-time rendering at over 200 FPS while consuming only 0.28 W for Large-Scale Real-World static scenes and 0.63 W for Large-Scale Real-World dynamic scenes, effectively addressing the challenge of deploying static/dynamic 3DGS on edge devices.

## ACKNOWLEDGMENTS

This work is supported in part by PRISM, one of the SRC/DARPA JUMP 2.0 centers, and by the Department of Health and Human Services Advanced Research Projects Agency for Health (ARPA-H) under Award Number AY1AX000003. The views and conclusions contained herein are those of the authors and should not be interpreted as necessarily representing the official policies or endorsements, either expressed or implied, of the Advanced Research Projects Agency for Health or the U.S. Government.